\documentclass[prb,twocolumn,showpacs,amsmath,amssymb,superscriptaddress,floatfix]{revtex4}

\usepackage{graphicx}%Include figure files
\usepackage{dcolumn}%Align table columns on decimal point
\usepackage{bm}% bold math
\usepackage{epsfig}

\begin{document}

\title{Spin and charge pumping in magnetic tunnel junctions with precessing magnetization: A nonequilibrium Green function approach}

\author{Son-Hsien Chen}
\email{d92222006@ntu.edu.tw}
\affiliation{Department of Physics and Astronomy, University
of Delaware, Newark, DE 19716-2570, USA}
\affiliation{Department of Physics, National Taiwan University, Taipei 10617, Taiwan}
\author{Ching-Ray Chang}
\email{crchang@phys.ntu.edu.tw}
\affiliation{Department of Physics, National Taiwan University, Taipei 10617, Taiwan}
\author{John Q. Xiao}
\affiliation{Department of Physics and Astronomy, University
of Delaware, Newark, DE 19716-2570, USA}
\author{Branislav K. Nikoli\' c}
\affiliation{Department of Physics and Astronomy, University
of Delaware, Newark, DE 19716-2570, USA}

\begin{abstract}
We study spin and charge currents pumped by precessing magnetization of a single ferromagnetic layer within $F|I|N$ or $F|I|F$ ($F$-ferromagnet; $I$-insulator; $N$-normal-metal) multilayers of nanoscale thickness attached to two normal metal electrodes with no applied bias voltage between them. Both simple one-dimensional model, consisting of a single precessing spin and a potential barrier as the ``sample,'' and realistic three-dimensional devices are investigated. In the rotating reference frame, where the magnetization appears to be static, these junctions are mapped onto a four-terminal dc circuit whose effectively half-metallic ferromagnetic electrodes are biased by the frequency $\hbar \omega/e$ of  microwave radiation driving magnetization precession at the ferromagnetic resonance (FMR) conditions. We show that pumped spin current in $F|I|F$ junctions, diminished behind the tunnel barrier and increased in the opposite direction, is filtered into charge current by the second $F$ layer to generate dc pumping voltage of the order of $\sim 1$ $\mu$V (at FMR frequency $\sim 10$ GHz) in an open circuit. In $F|I|N$ devices, several orders of magnitude smaller charge current and the corresponding dc voltage appear concomitantly with the pumped spin current due to barrier induced asymmetry in the transmission coefficients connecting the four electrodes in the rotating frame picture of pumping.
\end{abstract}

\maketitle

\section{Introduction}\label{sec:intro}

The pursuit of ``second generation'' spintronic devices~\cite{Awschalom2007} has largely been focused  on harnessing coherent spin states and their dynamics in metals and semiconductors. This requires to maintain and control spin orientations transverse to externally applied or internal magnetic fields. The salient example of phenomena involving both coherent spins and their time evolution is the {\em spin-transfer torque} where spin current of large enough density injected into a ferromagnetic layer either switches its magnetization from one static configuration to another or generates a dynamical situation with steady-state precessing magnetization.~\cite{Ralph2008} In the reciprocal effect, termed {\em spin pumping} because it occurs in setups without applied bias voltage,~\cite{Tserkovnyak2005} microwave driven precessing magnetization  of a single ferromagnetic layer under the FMR conditions emits {\em pure} spin current (not accompanied by any net charge flux~\cite{Nagaosa2008}) into adjacent normal-metal layers. In the conventional picture of spin pumping,~\cite{Tserkovnyak2005} $F|N$ interface pumps spin current in both directions,~\cite{Watts2006} so that its magnitude is determined by the interfacial parameters which govern transport of spins that are noncollinear to the magnetization direction at the interface.~\cite{Ralph2008,Tserkovnyak2005}

The  spin current emitted from the $F$ layer with moving magnetization has been observed~\cite{Mizukami2002,Gerrits2006,Heinrich2007} in early experiments only {\em indirectly} as an enhancement of Gilbert damping~\cite{Heinrich2007} of magnetization dynamics in inhomogeneous structures due to the presence of $F|N$ interfaces and fast relaxation of pumped spins in good ``spin sink''~\cite{Tserkovnyak2005} $N$ layers which ultimately leads to a loss of the angular momentum.~\cite{Taniguchi2007} Very recently it has been converted~\cite{Saitoh2006} into the conventionally measurable voltage signals through the inverse spin Hall effect (where longitudinal spin current injected into a metal with spin-orbit couplings generates transverse voltage between lateral edges of the sample~\cite{Nagaosa2008}). Another electrical scheme is based on $N_1|F|N_2$ multilayers~\cite{Costache2006} where different voltages develop at different $F|N_i$ interfaces due to backflow spin current (driven by the spin accumulation in $N_i$ layers built up by directly pumped spin current), which is detected by the precessing $F$ layer itself.~\cite{Wang2006g} These experiments suggest that spin pumping devices could be exploited as generators~\cite{Tserkovnyak2005,Gerrits2006} of elusive pure spin currents,~\cite{Nagaosa2008} where spin current injected from $F$ into adjacent  $N$ layers carries fast precessing spins in gigahertz range of frequencies  offering new functionality for metal spintronics.~\cite{Heinrich2007} They can also be used to probe important aspects of spin dynamics in thin $F$ layers.~\cite{Taniguchi2008}

Unlike spin-transfer torque that has been demonstrated in $F|I|F$ magnetic tunnel junctions (MTJ),~\cite{Ralph2008} it has been considered that low transparent interfaces would completely screen the interfacial spin pumping effect (as observed in some experiments~\cite{Lagae2005}), unless the tunnel barrier has non-trivial magnetic properties.~\cite{Xia2002} Thus, recent surprising measurements~\cite{Moriyama2008,Moriyama2009} of large voltage signals of the order of $\sim 1$ $\mu$V (at FMR frequencies $f \simeq 2$  GHz and precession cone angles $\theta \simeq 10^\circ$) in microwave driven $F|I|N$ and $F|I|F$ tunnel junctions, as opposed to $\sim 10$ nV pumping signals~\cite{Costache2006}  in $N_1|F|N_2$ multilayers, have attracted considerable theoretical attention.~\cite{Chui2008,Xiao2008,Tserkovnyak2008}  Nevertheless, {\em the puzzle of unexpectedly large magnitude} of the observed dc pumping voltages persists: ({\em i}) the scattering~\cite{Xiao2008}  approach to transport of noninteracting quasiparticles through defect-free epitaxial $F|I|F$ MTJ finds $\sim$ 1 nV signals at FMR frequency $f=2$  GHz and precession cone angle $\theta = 10^\circ$; ({\em ii}) the  tunneling Hamiltonian approach~\cite{Tserkovnyak2008} for clean $F|I|F$ MTJ and the same $f$ and $\theta$ parameters  sets the maximum dc pumping voltage at $\sim 0.01$ $\mu$V in parallel and $\sim 1$ $\mu$V in antiparallel configuration of two $F$ electrodes; and ({\em iii}) the tunneling Hamiltonian approach combined with semiclassical modeling of the interplay of spin diffusion and self-consistent screening around interfaces in $F|I|F$ and $F|I|N$ junctions involves too many unknown phenomenological parameters, thereby offering only a wide range of possible pumping voltages for both of these junctions.~\cite{Chui2008}

Here we address the problem of spin pumping and induced voltages by high frequency magnetization dynamics  in $F|I|F$ and $F|I|N$ junctions within the framework of nonequilibrium Green function (NEGF) formalism.~\cite{Haug2007,Nikoli'c2006} We note that NEGFs have been utilized before to study spin~\cite{Wang2003a,Hattori2007} and charge~\cite{Arrachea2006} pumping by time-dependent fields acting on finite-size paramagnetic devices attached to electrodes held at the same electrochemical potential. Since NEGF formalism takes as an input a microscopic Hamiltonian, it makes it possible to include, in a controlled fashion, the full geometry~\cite{Taniguchi2008a} of experimental devices (such as the finite thickness of $F$, $I$, and $N$ layers, down to few atomic monolayers, which play an important role in the magnetoresistance~\cite{Heiliger2007} and spin-transfer torque~\cite{Heiliger2008} of crystalline MTJs), the properties of the insulating barrier  (including disorder effects), as well as the interactions responsible for spin-flip processes in $F$. The NEGF formalism also makes it easy to take into account {\em ab initio} input~\cite{Heiliger2007,Heiliger2008,Heiliger2008b,Waldron2007} on the $F|I$ interface electronic and magnetic structure and the self-consistently developed nonequilibrium spin and charge distributions around it.

\begin{figure}
\centerline{\psfig{file=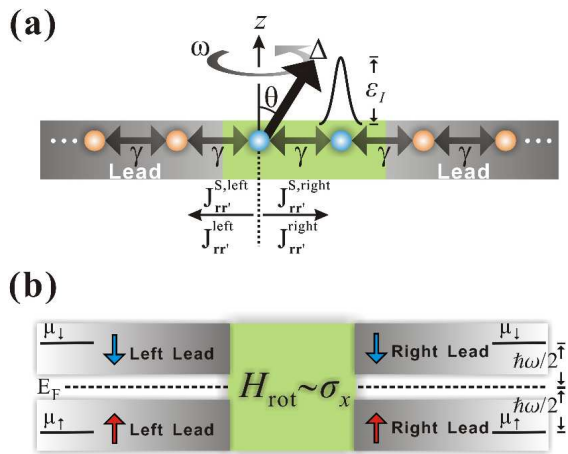,scale=0.35,angle=0}}
\caption{(color online). (a) The 1D model of spin pumping where the sample consisting of two sites, one hosting the single spin rotating with frequency
$\omega$ and the other one hosting the potential barrier of height $\varepsilon_I$, is attached to two semi-infinite tight-binding chains
($\gamma$ is the hopping parameter) playing the role of electrodes with no applied bias voltage between them. In the rotating reference frame the spin is static and the device
(a) is mapped into the four-terminal dc circuit in panel (b) whose electrodes have electrochemical potential shifted by $\pm \hbar \omega/2$ with respect to the equilibrium
Fermi level $E_F$ of unbiased electrodes in the laboratory frame.}
\label{fig:1d}
\end{figure}

Furthermore, NEGF approach yields a remarkably transparent physical picture of pumping in ferromagnetic multilayered systems. For example, in the simplest model of pumping, generated by a {\em single spin precessing} with frequency $\omega$ in Fig.~\ref{fig:1d}(a), the  NEGF rotated into the rest frame of the spin maps the original laboratory-frame device onto a dc circuit in  Fig.~\ref{fig:1d}(b). The central sample of this circuit, which contains time-independent spin interactions, is attached to four electrodes that allow only one spin species to propagate through them and are, therefore, labeled by $L$-left, $R$-right, spin-$\uparrow$, and spin-$\downarrow$. These four electrodes are biased by the voltage $\hbar \omega/e$, so that spin-$\downarrow$ electrons  flow from  electrodes at higher electrochemical potential and precess inside the sample due to spin-dependent interactions to be able to enter into electrodes at a lower electrochemical potential as spin-$\uparrow$ states. Thus, this picture reduces the quantitative analysis of spin and charge pumping by precessing spins to multiterminal Landauer-B\" uttiker-type formulas for spin-resolved charge currents as encountered in, e.g., the mesoscopic spin Hall effect.~\cite{Nikoli'c2005b,Nikoli'c2006}

The paper is organized as follows. In Sec.~\ref{sec:1d} we exploit the physical  picture of pumping  provided by Fig.~\ref{fig:1d}(b) to analyze {\em local} spin and charge currents flowing away from the single precessing spin toward the neighboring sites along the tight-binding chain in one-dimension (1D). This framework is extended to {\em total} pumped currents and associated  voltages in  three-dimensional (3D) multilayered structures, such as $F|N|F$, $F|I|F$, and $F|I|N$, in Sec.~\ref{sec:3d}. We conclude in Sec.~\ref{sec:conclusions}. Our principal results---pumped spin and charge currents in 1D model and voltage signals in $F|I|F$ and $F|I|N$ junctions---are shown in Figs.~\ref{fig:1d_currents} and \ref{fig:fif}, respectively.

\section{NEGF approach to spin and charge pumping by a single precessing spin in one dimension}\label{sec:1d}

The toy 1D model in Fig.~\ref{fig:1d}(a) encodes most of the essential physics of pumping by precessing spins while making it possible to obtain analytical solution for
the magnitude of pumped currents. For simplicity, we start from the often employed in spin-transfer torque~\cite{Theodonis2006} and spin pumping~\cite{Tserkovnyak2008} studies Stoner-type Hamiltonian~\cite{Fazekas1999}
\begin{eqnarray}\label{eq:hlab}
\hat{H}_{\rm lab}(t) & = &  \sum_{{\bf r}, \sigma, \sigma'} \left( \varepsilon_{\bf r}\delta_{\sigma\sigma'} - \frac{\Delta_{\bf r}}{2} \mathbf{m}_{\bf r}(t) \cdot \hat{\bm \sigma}^{\sigma\sigma'} \right) \hat{c}_{{\bf
r}\sigma}^\dag\hat{c}_{{\bf r}\sigma'} \nonumber \\
\displaystyle & & - \gamma \sum_{\langle {\bf rr'} \rangle \sigma} \hat{c}_{{\bf r}\sigma}^\dag \hat{c}_{{\bf r'}\sigma},
\end{eqnarray}
in the local orbital basis suited for NEGF calculations.~\cite{Haug2007,Nikoli'c2006} Its time dependence stems from the unit vector ${\bf m}(t)$ along the local magnetization direction, which is assumed to be spatially uniform and steadily precessing around the $z$ axis with a constant cone angle $\theta$. The operators $\hat{c}_{{\bf r}\sigma}^\dag$ ($\hat{c}_{{\bf r}\sigma}$) create (annihilate) electron with spin $\sigma$ at site ${\bf r}$, and $\gamma$ is the nearest neighbor hopping. The coupling of itinerant electrons to collective magnetic dynamics is described through the material-dependent exchange potential $\Delta_{\bf r}$, where $\hat{\bm \sigma}=(\hat{\sigma}_x,\hat{\sigma}_y,\hat{\sigma}_z)$ is the vector of the Pauli matrices and $\hat{\sigma}^{\sigma\sigma'}_i$ denotes the Pauli matrix elements. The on-site potential $\varepsilon_{\bf r}$ accounts for the presence of the barrier [such as $\varepsilon_{\bf r}=\varepsilon_I$ on the second site of the sample in Fig.~\ref{fig:1d}(a)], disorder, external electric field, and it can also be used to shift the band bottom conveniently. The sample is attached to two semi-infinite ideal (spin and charge interaction free) electrodes, which terminate in macroscopic reservoirs held at the same electrochemical potential $\mu_p=E_F$ where $E_F$ is the Fermi energy.

\begin{figure}
\centerline{\psfig{file=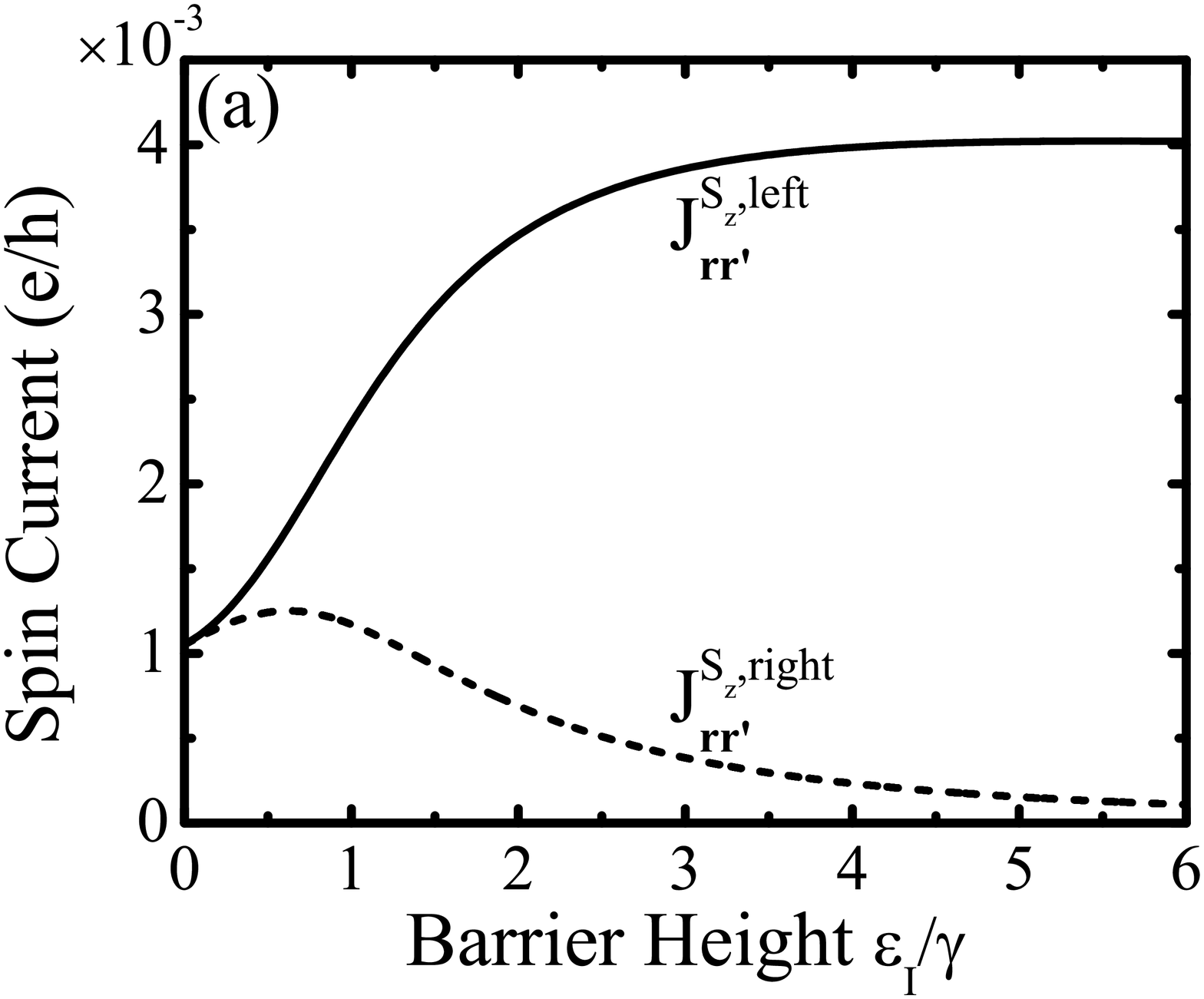,scale=0.25,angle=0}}
\centerline{\psfig{file=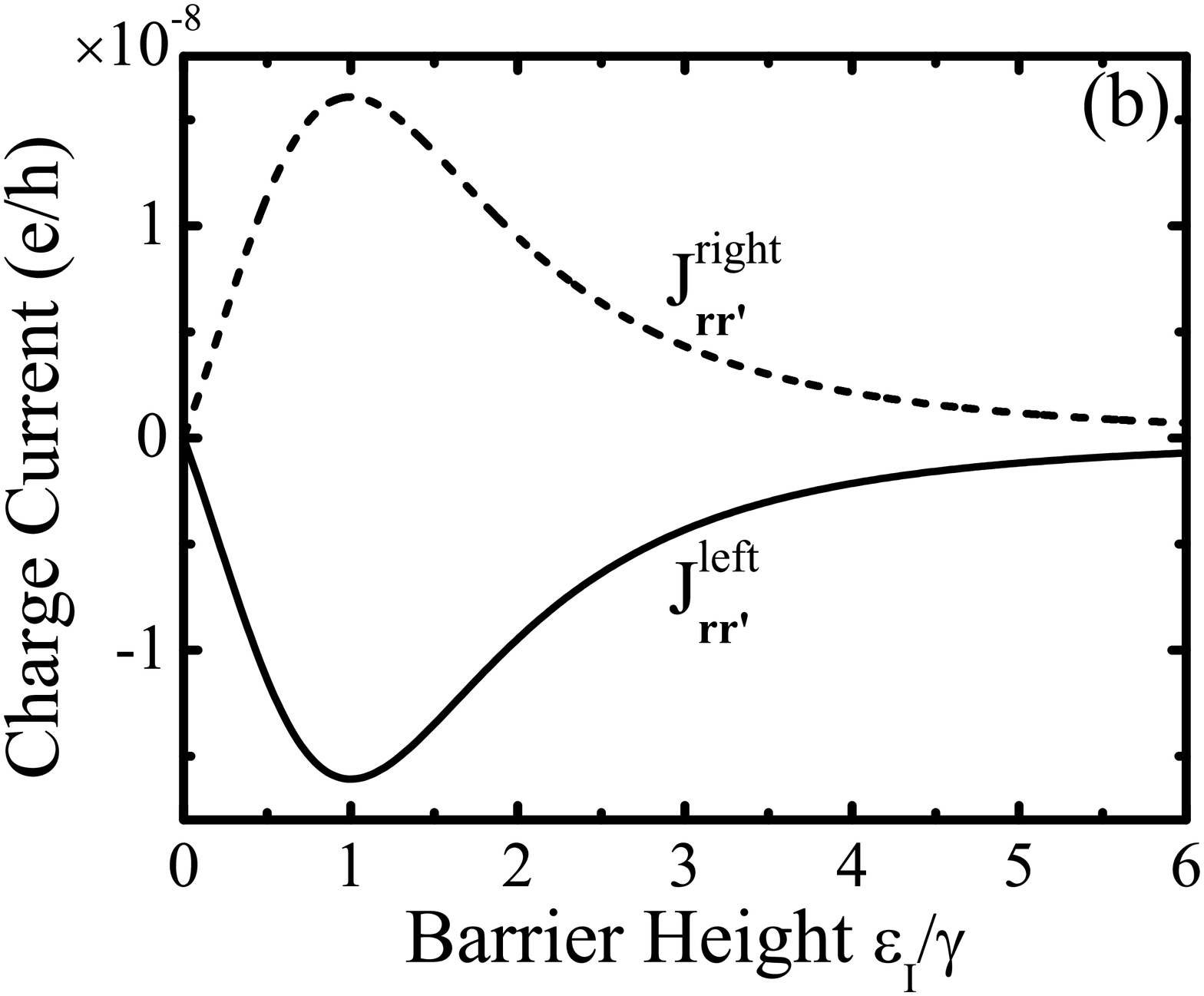,scale=0.25,angle=0}}
\caption{The (a) $S_z$-spin and (b) charge currents pumped by a single precessing spin as a function
of the potential barrier on the second site of the sample in 1D model shown in Fig.~\ref{fig:1d}(a). The parameters of the model
are: $f = \omega/2\pi = 20$ GHz; $\theta=10^\circ$, $\Delta/E_F = 0.85$, and electrons in the macroscopic reservoirs to which the electrodes are attached
have the Fermi energy $E_F=2 \gamma$.}
\label{fig:1d_currents}
\end{figure}

The fundamental objects~\cite{Haug2007} of the NEGF formalism are the retarded
\begin{equation}\label{eq:retarded}
G^{r,\sigma\sigma'}_{\bf rr'}(t,t')=-\frac{i}{\hbar} \Theta(t-t') \langle \{\hat{c}_{{\bf r}\sigma}(t) , \hat{c}^\dagger_{{\bf r'}\sigma'}(t')\}\rangle,
\end{equation}
and the lesser
\begin{equation}\label{eq:lesser}
G^{<,\sigma\sigma'}_{\bf rr'}(t,t')=\frac{i}{\hbar} \langle \hat{c}^\dagger_{{\bf r'}\sigma'}(t') \hat{c}_{{\bf r}\sigma}(t)\rangle,
\end{equation}
Green functions ($\langle \ldots \rangle$ denotes the nonequilibrium statistical average~\cite{Haug2007}) which describe the density of available quantum states and how electrons occupy those states, respectively.  Since nonequilibrium problems are not time-translation invariant, these Green functions depend on two time variables separately. However, the
cumbersome double time dependence of NEGF in general pumping problems~\cite{Arrachea2006} can be eliminated~\cite{Hattori2007} for the special case of time-dependent potential caused by precessing magnetization using the compensating rotation~\cite{Ballentine1998} of the system described by the unitary transformation $\hat{U}=e^{i \omega \hat{\sigma}_z t/2}$ (for magnetization precessing counterclockwise). Thus, the Hamiltonian in the rotating frame~\cite{Tserkovnyak2008}
\begin{equation}\label{eq:hrot}
\hat{H}_{\rm rot}  =  \hat{U} \hat{H}_{\rm lab}(t) \hat{U}^{\dagger} - i \hbar \hat{U} \frac{\partial}{\partial t} \hat{U}^\dagger = \hat{H}_{\rm lab}(0)  - \frac{\hbar \omega}{2} \hat{\sigma}_z,
\end{equation}
is time-independent. The term $\hbar \omega\hat{\sigma}_z/2$, which appears uniformly in the Hamiltonian of the sample or $N$ electrodes, will spin-split the bands
of the $N$ electrodes. This yields a rotating-frame picture of pumping based on the four-terminal device in Fig.~\ref{fig:1d}(b).

The device in Fig.~\ref{fig:1d}(b) guides us in setting up the NEGF equations for the description of currents flowing between its four electrodes, labeled by $p,\sigma$ ($p=L,R$ and $\sigma=\uparrow,\downarrow$), which are biased by the voltage $\hbar \omega/e$. The electrodes behave effectively as the half-metallic ferromagnets, emitting or absorbing only one spin species. The rotating frame Green functions
\begin{equation}
\bar{\bf G}^r(E)=\left[E-\bar{\bf H}_{\rm rot}-\bar{\bf \Sigma}^r(E)\right]^{-1},
\end{equation}
and
\begin{equation}
\bar{\bf G}^<(E) = \bar{\bf G}^r(E) \bar{\bf \Sigma}^<(E) \bar{\bf G}^a(E),
\end{equation}
depend on $\tau=t-t^\prime$, or energy $E$ after the time difference $\tau$ is Fourier transformed. Here the advanced Green function is $\bar{\bf G}^a(E)=[\bar{\bf G}^r(E)]^\dagger$, and $\bar{\bf H}_{\rm rot}$ is the matrix representing $\hat{H}_{\rm rot}$ in the local-orbital basis. The retarded self-energy matrix $\bar{\bf \Sigma}^r(E)=\sum_{p,\sigma} \bar{\bf \Sigma}^{r,\sigma}_p(E)$ is the sum of self-energies introduced by the interaction with the leads which determine escape rates of spin-$\sigma$ electron into the electrodes $p,\sigma$ in Fig.~\ref{fig:1d}(b).

For interacting systems $\bar{\bf \Sigma}^r(E)$ would also contain electron-electron and electron-phonon contributions, while for noninteracting systems, described by Hamiltonian (\ref{eq:hrot}), the lesser self-energy is expressed in terms of $\bar{\bf \Sigma}^{r,\sigma}_p(E)$ as
\begin{equation}\label{eq:lesser_se}
\bar{\bf \Sigma}^<(E)=\sum_{p,\sigma} i f_p^{\sigma}(E) \bar{\bf \Gamma}_p^{\sigma}(E).
\end{equation}
The level broadening matrix
\begin{equation}
\bar{\bf \Gamma}_p^\sigma(E) = -2 {\rm Im}\, {\bf \Sigma}^r_p \left(E + \sigma \frac{\hbar \omega}{2} \right),
\end{equation}
is obtained from the usual self-energy matrices~\cite{Haug2007} ${\bf \Sigma}^r_p(E)$ of semi-infinite leads in the laboratory frame with their energy argument being shifted by $\sigma  \hbar \omega/2$ to take into account the ``bias voltage'' in accord with Fig.~\ref{fig:1d}(b). The distribution function of electrons in the four electrodes of the rotating frame dc circuit is given by
\begin{equation}\label{eq:fermi}
f_p^{\sigma}(E)=\frac{1}{\exp[(E-E_F+\sigma\hbar\omega/2)/kT]+1},
\end{equation}
where $\sigma=+$ for spin-$\uparrow$ and $\sigma=-$ for spin-$\downarrow$. Since the device is not biased in the laboratory frame, the shifted Fermi function in Eq.~(\ref{eq:fermi}) is uniquely specified by the polarization $\uparrow$ or $\downarrow$ of the  electrode, so that we remove the lead label $p$ from it in the equations below.

The basic transport quantity for the rotating-frame dc circuit is the spin-resolved bond charge current~\cite{Nikoli'c2006} carrying spin-$\sigma$ electrons from site ${\bf r}$ to neighboring site ${\bf r}'$
\begin{equation}\label{eq:jlab}
J_{\bf r r'}^\sigma = \frac{e\gamma}{h} \int\limits_{-\infty}^{\infty} dE\, [\bar{G}^{<,\sigma \sigma}_{\bf r'r}(E) - \bar{G}^{<,\sigma\sigma}_{\bf rr'}(E)].
\end{equation}
This gives spin
\begin{equation}
J^S_{\bf rr'}=J^{\uparrow}_{\bf rr'}-J^{\downarrow}_{\bf rr'},
\end{equation}
and charge
\begin{equation}
J_{\bf rr'}=J^{\uparrow}_{\bf rr'}+J^{\downarrow}_{\bf rr'},
\end{equation}
bond currents flowing between neighboring sites.~\cite{Nikoli'c2006} Equation~(\ref{eq:jlab}) can be evaluated analytically for 1D model assuming that for small enough $\hbar\omega \ll E_F$ we can use $f^\downarrow(E)-f^\uparrow(E)=\hbar \omega \delta(E-E_F)$ at zero temperature. Such ``adiabatic approximation''~\cite{Hattori2007} is analogous to linear-response limit in conventional transport calculations for devices biased by small voltage difference.

The $S_z$-component of the bond spin current between the precessing site and its nearest neighbor in the sample, as illustrated in Fig.~\ref{fig:1d}(a), is given by
\begin{eqnarray}\label{eq:spinz_current}
J_{\bf rr'}^{S_z} & = & \hbar \omega \sin^2 \theta \frac{\gamma^{2} \Delta^{2}({\rm Im}\, \Sigma_{\rm 1D})^2}{8\pi |R|^2} \nonumber \\
\displaystyle && \times  \left[ 4\left(\gamma^{2}+\varepsilon_{I} ^{2}\right) + 4  | \Sigma_{\rm 1D}|^2 - 8 \varepsilon_{I} {\rm Re}\, \Sigma_{\rm 1D} \right],
\end{eqnarray}
where the terms $O \left[ (\hbar\omega)^{2} \right]$ are neglected. Here $\Sigma_{\rm 1D}= ( E_F-2\gamma - \sqrt{(E_F-2\gamma)^2-4\gamma^2})/2$ is the self-energy of 1D semi-infinite  lead (i.e., tight-binding chain) and $R = (4 \Sigma_{\rm 1D}^ 2- \Delta^{2})(\Sigma_{\rm 1D} - \varepsilon_{I})^2/4 +(4 \varepsilon_I \Sigma_{\rm 1D}-4\Sigma_{\rm 1D}^2) \gamma^2/2 + \gamma^4$. In deriving Eq.~(\ref{eq:spinz_current}), we assume uniform band bottom, so that  $\varepsilon _{I} \mapsto \varepsilon_{I} - (\Delta \cos \theta)/2$ and $E_{F} \mapsto E_{F} + (\Delta \cos \theta)/2$ is used to plot Fig.~\ref{fig:1d}.

The expression in Eq.~(\ref{eq:spinz_current}) reproduces all major features of the scattering approach~\cite{Tserkovnyak2005} to adiabatic ($\hbar \omega \ll \Delta$) regime of spin pumping by $F|N$ interface in 3D multilayers: ({\em i})  the pure spin current carrying $S_z$ spins is proportional to $\hbar \omega$ and $\sin^2 \theta$; ({\em ii})  $S_z$ component $J_{\bf rr'}^{S_z}$ of the spin current tensor is time independent in both rotating and laboratory frames; and ({\em iii}) $S_x$- and $S_y$-components of the pumped spin current oscillate harmonically with time in the laboratory frame. Moreover, when potential barrier $\varepsilon_I$ is introduced into the sample, we find in Fig.~\ref{fig:1d_currents}(a) that spin current on the right decays with increasing $\varepsilon_I$ while pumped spin current flowing on the left increases to about twice the value of the sum $J_{\bf rr'}^{S_z,{\rm left}} + J_{\bf rr'}^{S_z,{\rm right}}$ of the left and right spin currents pumped symmetrically in the absence of the barrier. This effect can clarify the origin of possible Gilbert damping enhancement in realistic MTJ devices consisting of $N|F|I|F|N$ multilayers, rather than infinite $F$ electrodes, where angular-momentum loss develops due to increasing spin pumping into the left $N$ electrode even when the insulating barrier $I$ suppresses spin pumping on the right side of the junction.

Figure ~\ref{fig:1d_currents}(b) demonstrates that non-zero potential $\varepsilon_I \neq 0$ also leads to concomitant pumping of a tiny  charge current into the right electrode, which is several order of magnitude smaller than the pumped spin current. Unlike pumping of spins which is linear in frequency, such pumped charge current scales as  $\sim \omega^2$. We discuss its origin in Sec.~\ref{sec:3d} by analyzing total charge current in the $N$ terminals expressed in terms of the transmission coefficients between the fully spin-polarized electrodes in the rotating frame.

\section{NEGF approach to spin and charge pumping in 3D $\bm{F|N|F}$, $\bm{F|I|F}$, and $\bm{F|I|N}$ multilayers}\label{sec:3d}

\begin{figure}
\centerline{\psfig{file=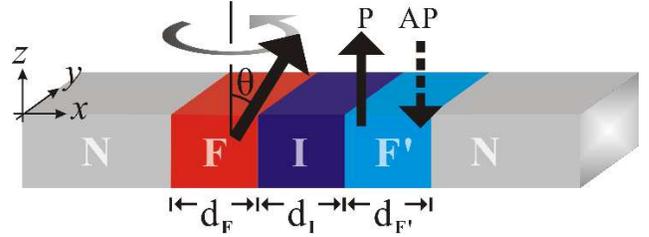,scale=0.45,angle=0}}
\caption{(color online). The magnetic tunnel junction with precessing magnetization in the left $F$ layer is modeled on a simple-cubic tight-binding lattice. The thicknesses of the ferromagnetic ($F$, $F'$) and thin insulating ($I$) layers are measured using the number of atomic monolayers $d_F$, $d_{F'}$, and $d_I$, respectively. The P and AP configurations of MTJ correspond to the magnetization of the right $F$ layer being parallel or antiparallel to the $z$-axis around which spatially uniform magnetization of the left $F$ layer steadily precesses with a constant cone angle $\theta$.}
\label{fig:3d}
\end{figure}

We extend this analysis to a 3D MTJ shown in Fig.~\ref{fig:3d} which consists~\cite{Theodonis2006} of infinite planes of $F$, $N$, and $I$ materials modeled on a simple-cubic tight-binding lattice with single $s$-orbital per site using Hamiltonian (\ref{eq:hlab}). The effective dc circuit [Fig.~\ref{fig:1d}(b)] in the rotating frame makes it easy to write the expression for the total charge current in the left and right $N$ electrodes. For example, spin-$\downarrow$ electrons can flow from  $_L^\downarrow$ lead at higher electrochemical potential into $_R^\uparrow$ lead at the lower electrochemical potential. They enter $_R^\uparrow$ lead as spin-$\uparrow$ electrons with probability determined by the precession inside the sample since $\hat{H}_{\rm rot}$ contains terms proportional to $\hat{\sigma}_x$ for which the injected spin states $|\!\! \downarrow\rangle$ from $_L^\downarrow$ lead (polarized along the $z$ axis) are not the eigenstates.

Since pumped charge current is necessarily conserved, as exemplified by Fig.~\ref{fig:1d_currents}(b), we arbitrarily select the right electrode (current flowing into the electrode is assumed to be positive) to find its explicit expression in terms of the multiterminal Landauer-B\" uttiker formulas~\cite{Nikoli'c2005b} for spin-resolved quantum transport:
\begin{eqnarray}\label{eq:total_charge}
I & = & \frac{e}{h} \int\limits_{-\infty}^\infty\!\!\! dE \, \left\{T_{RL}^{\uparrow \downarrow}[f^\downarrow(E)-f^\uparrow(E)] \right. \nonumber \\
\displaystyle && - \left. T_{LR}^{\uparrow \downarrow}[f^\downarrow(E)-f^\uparrow(E)] \right\}.
\end{eqnarray}
Here the transmission coefficients $T_{pp'}^{\sigma \sigma'}$ determine the probability for $\sigma'$ electrons injected through lead $p'$ to emerge in electrode $p$ as spin-$\sigma$ electrons. They can be computed from the NEGF-based formula~\cite{Haug2007}
\begin{equation}\label{eq:caroli}
T_{pp'}^{\sigma \sigma'} = {\rm Tr}\, \left\{\bar{\bf \Gamma}_p^\sigma \bar{\bf G}^{r,\sigma\sigma'}_{pp'} \bar{\bf \Gamma}_{p'}^{\sigma'} [\bar{\bf G}^{r,\sigma\sigma'}_{pp'}]^\dagger\right\},
\end{equation}
which is written here in the spin-resolved form. The block $\bar{\bf G}_{p p'}^{r,\sigma \sigma^\prime}$ of the retarded Green function matrix consists of those matrix elements which connect the layer of the sample attached to lead $p'$ to the layer of the sample attached to lead $p$.

In general, the spin current is not conserved, as illustrated by  Fig.~\ref{fig:1d_currents}(a), and we choose to compute it in the left $N$ electrode:
\begin{eqnarray}\label{eq:total_spin}
I^{S}_L & = &  \frac{e}{h} \int\limits_{-\infty}^\infty\!\!\! dE \, \left\{T_{LR}^{\uparrow \downarrow}[f^\downarrow(E)-f^\uparrow(E)] \right. \nonumber \\
\displaystyle && +  T_{RL}^{\uparrow \downarrow}[f^\downarrow(E)-f^\uparrow(E)]  \nonumber \\
\displaystyle && + \left. 2T_{LL}^{\uparrow \downarrow} [f^\downarrow(E)-f^\uparrow(E)] \right\}.
\end{eqnarray}
 The expressions Eq.~(\ref{eq:total_charge}) and Eq.~(\ref{eq:total_spin}) for total currents are equivalent to the sum of all bond charge $J_{\bf rr'}$ or bond spin  $J^S_{\bf rr'}$ currents, respectively, where summation is performed over the pairs of sites within the electrode at a chosen cross section.~\cite{Nikoli'c2006} By the same token, the analytical expression Eq.~(\ref{eq:spinz_current}) for the spin current is already equivalent to the result obtained from Eq.~(\ref{eq:total_spin}) since no summation is necessary for the cross section consisting of a single site.

\begin{figure}
\centerline{\psfig{file=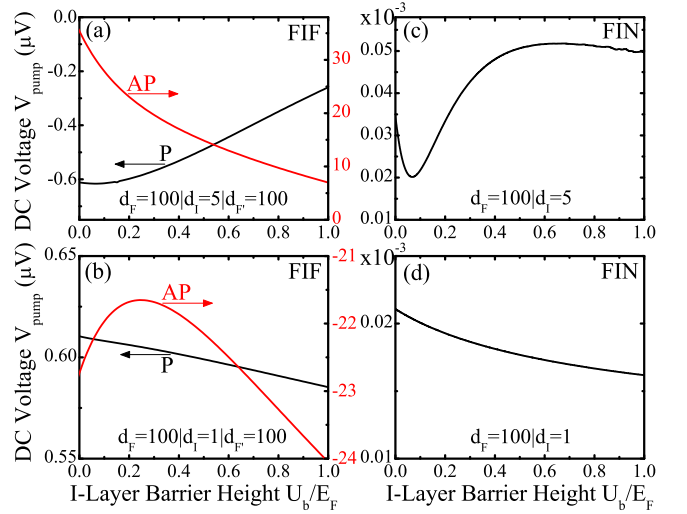,scale=0.32,angle=0}}
\caption{(Color online). The dc pumping voltage in [(a) and (b)] $F|I|F$ and [(c) and (d)] $F|I|N$ multilayers attached to two semi-infinite $N$ electrodes as the function of the barrier height $U_b$ (measured relative to the Fermi energy). The magnetization of the left $F$ layer is precessing with cone angle $\theta=10^\circ$ at frequency $f=\omega/2\pi=20$ GHz. The parameters  describing the multilayer are $E_F=4.5$ eV, $\Delta/E_F=0.85$ (in both $F$
layers for $F|I|F$ junction), and $\gamma=1.0$ eV.}
\label{fig:fif}
\end{figure}

The pumped charge current in multilayers with the second analyzing $F$ layer originates from spin filtering by the static magnetization of the analyzing $F$ layer of current pumped toward the right. That is, we find $I^{S_z}_R =0$, $I^{S_z}_L \neq 0$ and $I \neq 0$ in such systems. In junctions with a single precessing $F$ layer the pumped spin current is {\em pure} if $I^{S_z}_L=I^{S_z}_R \neq 0$ and $I = 0$. The possibility of nonzero pumped charge current even in junctions with only one $F$ layer whose magnetization is precessing, as exemplified by Fig.~\ref{fig:1d_currents}(b) and $F|I|N$ junctions in general [see Figs.~\ref{fig:fif}(c) and \ref{fig:fif}(d)], is explained by Eq.~(\ref{eq:total_charge}) as the consequence of the asymmetry in transmission coefficients  $T_{RL}^{\uparrow \downarrow} - T_{LR}^{\uparrow \downarrow} \neq 0$ when arbitrary potential $\varepsilon_I \neq 0$ is introduced in one of the layers.

The transmission coefficients can also explain the unexpectedly large enhancement of $I^{S_z}_L$ or $J_{\bf m m'}^{S_z,{\rm left}}$ in Fig.~\ref{fig:1d_currents}. As the barrier height $\varepsilon_I$ increases, $T_{LR}^{\uparrow \downarrow}$ and  $T_{RL}^{\uparrow \downarrow}$ diminish to very small value while $2T_{LL}^{\uparrow \downarrow}$ increases to about four times its value at $\varepsilon_I=0$  due to quantum interferences effects on the left side of the device (quantum interferences were also found to enhance pumped spin current when coherent backscattering from disorder occurs in finite-size conductors at paramagnetic resonance~\cite{Hattori2007}). At $\varepsilon_I=0$, $T_{LR}^{\uparrow \downarrow} = T_{RL}^{\uparrow \downarrow} = T_{LL}^{\uparrow \downarrow}$ so that spin currents of the same magnitude are pumped in both directions symmetrically.

The pumped charge current is translated into dc voltage in open circuits via
\begin{equation}\label{eq:vpump}
V_{\rm pump}=\frac{I}{G(\theta)},
\end{equation}
where $G(\theta)$ is the conductance of $F|I|F$ (or $F|I|N$ when the second $F$ layers is removed) junction sketched in Fig.~\ref{fig:3d} whose first $F$ layer has its static magnetization tilted by an angle $\theta$ away from the $z$ axis and the linear response bias voltage is applied between the $N$ electrodes in the laboratory frame. The quantity $G(\theta)=2e^2T_{RL}/h$ can also be computed via the standard NEGF formula~\cite{Haug2007} as in Eq.~(\ref{eq:caroli}) but for total $T_{RL}$, rather than spin-resolved, transmission coefficient expressed in terms of the retarded Green function and self-energies in the laboratory frame.

The largest voltage signal of spin pumping is expected in high quality epitaxial ${\rm Fe}|{\rm MgO}|{\rm Fe}$ tunnel junctions.~\cite{Tserkovnyak2008} To mimic their huge tunneling magnetoresistance (TMR),  while using the simple single-orbital tight-binding Hamiltonian ~(\ref{eq:hlab}), we adopt the same parameters employed in Ref.~\onlinecite{Xiao2008}: $E_F=4.5$ eV, $\Delta/E_F=0.85$, $\gamma=1.0$ eV, and the barrier height measured relative to the Fermi energy $U_b=(\varepsilon_I-E_F)$ is $U_b/E_F=0.25$. The band bottom is aligned across all layers of the junction with the bottom of the band for majority spins in $F$ (similarly to Ref.~\onlinecite{Xiao2008}). The ``optimistic'' TMR  ratio for this junction with $d_I=5$ monolayers of the insulating material is ${\rm TMR}=(R_{\rm AP}-R_{\rm P})/R_{\rm P} \simeq 3900\%$, which is close to {\em ab initio} computed zero-bias ${\rm TMR} \simeq 3700\%$ for defect-free  ${\rm Fe}|{\rm MgO}|{\rm Fe}$ MTJ containing five MgO layers.~\cite{Waldron2007}

In the coherent limit of tunneling,~\cite{Waldron2007} applicable to ideal crystalline structures without any defect scattering, the in-plane wave vector ${\bf k}_{||}=(k_y,k_z)$ is conserved and
all NEGF quantities depend on it. This requires to integrate $T_{pp'}^{\sigma \sigma'}(E,{\bf k}_{||})$ in Eq.~(\ref{eq:total_charge}) and ~(\ref{eq:total_spin}) over the two-dimensional (2D) Brillouin zone (BZ). Thus, in the adiabatic limit and at zero temperature we use the following formulas to obtain the charge current
\begin{equation}
I  = \frac{e \omega}{2\pi} \int\limits_{\rm BZ} \! d{\bf k}_{||} \,  (T_{RL}^{\uparrow \downarrow}- T_{LR}^{\uparrow \downarrow}),
\end{equation}
and the spin current in the left lead
\begin{equation}
I^{S}_L  =  \frac{e\omega}{2\pi} \int\limits_{\rm BZ} \! d{\bf k}_{||} \, (T_{LR}^{\uparrow \downarrow} + T_{RL}^{\uparrow \downarrow} + 2T_{LL}^{\uparrow \downarrow}),
\end{equation}
pumped by magnetization precessing at frequency $\omega$.

\begin{figure}
\centerline{\psfig{file=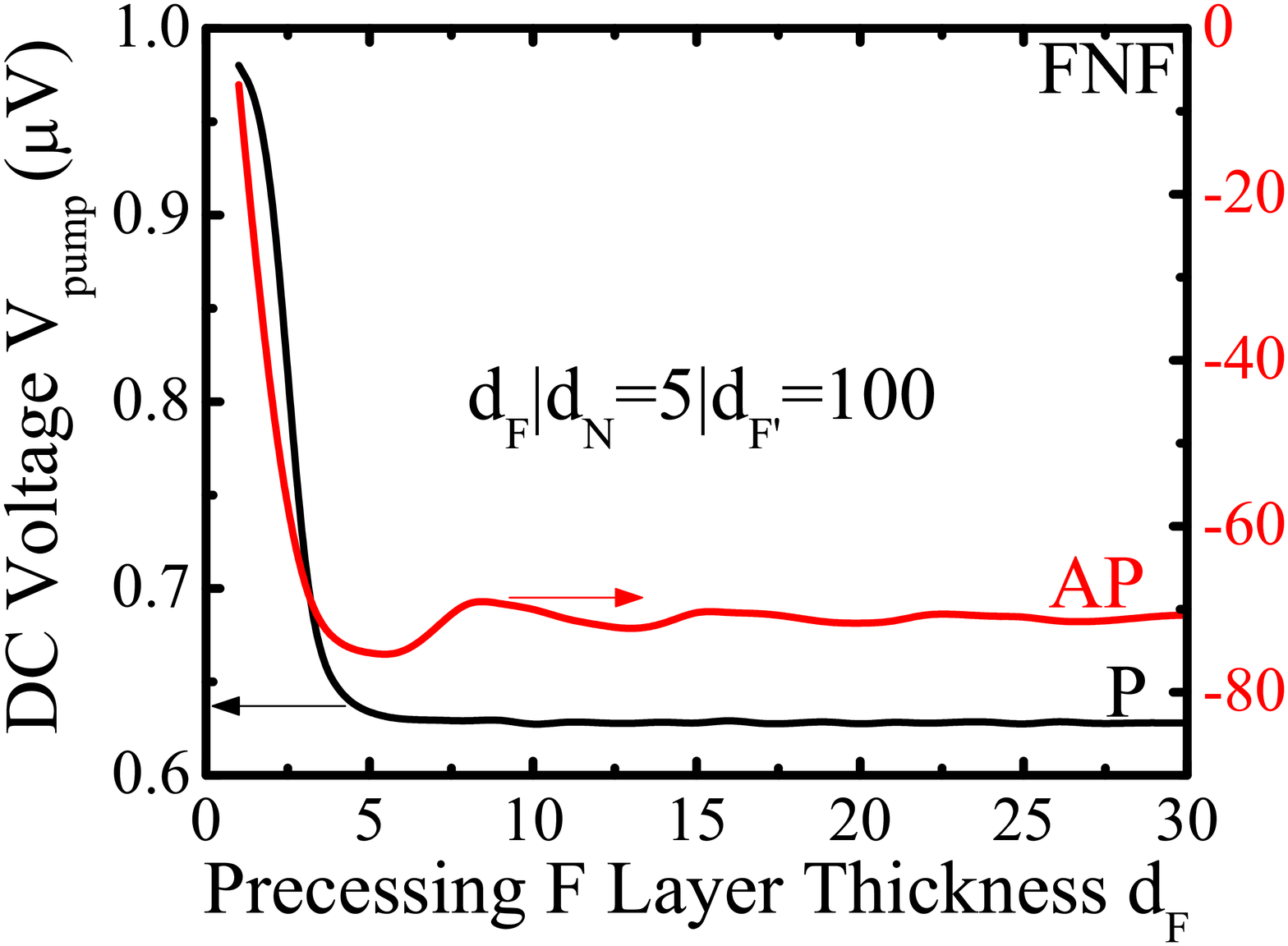,scale=0.25,angle=0}}
\caption{(color online). The dc pumping voltage in $F|N|F$ multilayers attached to two semi-infinite $N$ electrodes as the function of the thickness of $F$ layer whose magnetization is precessing with cone angle $\theta=10^\circ$ at frequency $f=\omega/2\pi =20$ GHz. The parameters describing the multilayer are $E_F=4.5$ eV, $\Delta/E_F=0.85$ (in both $F$ layers), and $\gamma=1.0$ eV.}
\label{fig:fnf}
\end{figure}

The computational algorithm for this integration can be {\em substantially} accelerated by transforming the 2D planar momentum integral
into a single integral over the in-plane kinetic energy
\begin{eqnarray}
\int\limits_{-\pi/a}^{\pi/a} \! \int\limits_{-\pi/a}^{\pi/a} dk_y dk_z \, T_{pp'}^{\sigma \sigma'}(E,{\bf k}_{||}) & = & \left(\frac{2\pi}{a}\right)^2 \! \int\limits_{-\infty}^\infty \! d\varepsilon_{yz} \, \rho_{\rm 2D}(\varepsilon_{yz}) \nonumber \\
\displaystyle && \times  T_{pp'}^{\sigma \sigma'}(E,\varepsilon_{yz}),
\end{eqnarray}
where we utilize the two-dimensional density of states $\rho_{\rm 2D}(\varepsilon_{yz})$ for a square lattice and the fact that $T_{pp'}^{\sigma \sigma'}$ depends on ${\bf k}_{||}$ through the in-plane kinetic energy $\varepsilon_{yz}$. In the case of nearest-neighbor hopping on a square lattice, the kinetic energy within a monolayer is given by $\varepsilon_{yz}=4\gamma-2\gamma[\cos(k_ya)+\cos(k_za)]$, where $a$ is the lattice spacing. The effect of the in-plane kinetic energy is equivalent to an increase in the on-site potential $\varepsilon_{\bf r} \mapsto \varepsilon_{\bf r} + \varepsilon_{yz}$.

To provide reference values for understanding the magnitude of pumping voltages in tunnel junctions, as well as to connect our theory to a ``standard model'' of interfacial spin pumping provided by the scattering theory,~\cite{Tserkovnyak2005,Watts2006,Heinrich2007} we first compute the dc voltage $V_{\rm pump}$ generated in $F|N|F$ multilayers. The chosen cone angle $\theta=10^\circ$ and FMR frequency $f= \omega/2\pi =20$ GHz are within the range of typical values encountered in experiments~\cite{Costache2006} where the results in Figs.~\ref{fig:fnf} and ~\ref{fig:fif} can easily be rescaled for other values of these two parameters using the general $ \propto \sin^2 \theta$ and $\propto \omega$ dependence in Eq.~(\ref{eq:spinz_current}). Figure~\ref{fig:fnf} demonstrates that pumping involves only a thin layer of $F$ material around the $F|N$ interface. However, while in the scattering theory~\cite{Tserkovnyak2005} adiabatic pumping develops over the atomistically short ferromagnetic coherence length $\sim \hbar v_F/\Delta$, which in our junction is $\hbar v_F/\Delta \simeq a$, we find that pumping in Fig.~\ref{fig:fif} involves about five monolayers of the ferromagnetic material. Here we assume that the magnitude of pumped current generated on this length scale is not affected~\cite{Tserkovnyak2008} by spin-relaxation processes [not included in Hamiltonian (\ref{eq:hlab})] that typically occur on a much longer length scale.~\cite{Bass2007} The pumped voltages in both P (parallel) and AP (antiparallel) configuration are below the maximum~\cite{Tserkovnyak2005,Xiao2008} expected voltage $V_{\rm pump} < \hbar \omega \approx 83$ $\mu$V (for the explanation of P and AP junction setups in the context of pumping by precessing magnetization, see Fig.~\ref{fig:3d}).

The dc pumping voltage for tunnel junctions is shown in Fig.~\ref{fig:fif}. Although the presence of the potential barrier within $I$ layer of $F|I|F$ junction increases the resistance of the junction in Eq.~(\ref{eq:vpump}), the pumped charge current decreases faster so that $V_{\rm pump}$ decreases with increasing barrier height $U_b$. In contrast to the scattering result of Ref.~\onlinecite{Xiao2008} where $V_{\rm pump}$ increases with increasing $U_b$ for all thicknesses of the $I$ layer, we find in Fig.~\ref{fig:fif}(b) such increase only if the $I$ layer consists of a single monolayer. The large difference between $V_{\rm pump}^{\rm P}$ and $V_{\rm pump}^{\rm AP}$ configuration stems from huge TMR ratio for this junction while the magnitude of pumped charge current remains virtually the same for both P and AP configurations.

These results are quite close to $\sim 1$ $\mu$V for $V_{\rm pump}^{\rm AP} - V_{\rm pump}^{\rm P}$ observed in MTJs with Al$_2$O$_3$ barriers, and an order of magnitude larger voltages in MTJs with MgO barriers.~\cite{Moriyama2009} The experimentally observed~\cite{Moriyama2009} change in sign of $V_{\rm pump}$ depending on the type of the barrier (Al$_2$O$_3$ vs. MgO) or its thickness might be related to a difference in sign between Figs.~\ref{fig:fif}(a) and \ref{fig:fif}(b). We also compute the pumped spin currents in the left $I^{S_z}_L=0.012 e/h$ and right $I^{S_z}_R=0$ electrodes, which do not depend on $\varepsilon_I$ or $U_b$ in the range shown in Fig.~\ref{fig:fif}.

Analogously to pumped charge current of 1D model in Fig.~\ref{fig:1d_currents}(b), we find nonzero charge current and corresponding dc pumping voltage in
$F|I|N$ junctions shown in Figs.~\ref{fig:fif}(c) and \ref{fig:fif}(d). Nevertheless, $V_{\rm pump}$ of the order of $\sim 10$ pV are way to small to explain recent experiments
on $F|I|N$ junctions~\cite{Moriyama2008} where $V_{\rm pump} \simeq 1$ $\mu$V is measured at frequencies of the applied rf field in the range  $f=2$--$3$ GHz and
the precession cone angle $\theta=10^\circ$--$17^\circ$ tuned by the microwave input power.

\section{Concluding Remarks}\label{sec:conclusions}
In conclusion, we have demonstrated that pumping of spin and charge currents by the precessing magnetization of a ferromagnetic layer within various multilayer setups  consisting of $F$, $N$, and $I$ layers of nanoscale thicknesses can be understood within the framework of NEGF rotated into the frame moving with the magnetization as a simple four-terminal dc circuit problem, as illustrated by Fig.~\ref{fig:1d}(b). The four leads of this circuit  are labeled as: $_L^\uparrow$, $_L^\downarrow$, $_R^\uparrow$, and $_R^\downarrow$ (i.e., they act as half-metallic ferromagnetic electrodes). They are biased by the voltage difference $\hbar \omega/e$ effectively emerging between the electrodes of opposite polarization. Our formalism provides a transparent physical picture of how: ({\em i}) single precessing spin pumps {\em pure} spin current symmetrically (in the absence of any barriers) toward the left and the right in 1D; ({\em ii}) pumped spin currents are suppressed by the tunnel barrier in one direction and enhanced in the opposite direction beyond na\" ive sum of currents before the introduction of the barrier; ({\em iii}) pumped spin currents develop over few monolayers of $F$ material in 3D junctions; and ({\em iv}) pumped spin currents become filtered by the second $F$ layer with static magnetization which converts them into charge current and the corresponding dc pumping voltage in open circuits. Our physical picture of spin and charge pumping in MTJs with time-dependent  magnetization suggests that these setups can serve as a sensitive probe of MTJ parameters, such as the properties of the tunnel barrier and damping parameters.

The pumping voltages in $N|F|I|F|N$ tunnel junctions of the order of  $\sim 1$ $\mu$V at FMR frequencies $\sim 10$ GHz  could explain some of the recent measurements of large voltage signals in  microwave driven MTJs under the FMR conditions.~\cite{Moriyama2009} They are much larger than  $\sim 10$ nV signal (at FMR frequencies $\sim 10$ GHz) recently predicted by the scattering theory~\cite{Xiao2008} for MTJs with similar TMR, but whose infinite $F$ electrodes are assumed to have strong spin-flip scattering leading to a vanishing spin accumulation in $F$.~\cite{Xiao2008} The spin-flip scattering can easily be introduced in Hamiltonian (\ref{eq:hlab}) via spin-orbit (SO) coupling terms~\cite{Nikoli'c2007} whose strength is tuned to match experimental values for spin-diffusion length.~\cite{Bass2007} Nevertheless, here we use simpler Hamiltonian following   assumptions  similar to Ref.~\onlinecite{Tserkovnyak2008}---typical spin-relaxation lengths~\cite{Bass2007} are much longer than the length scale (illustrated by Fig.~\ref{fig:fnf}) over which pumping develops so that it does not affect the strength of pumped currents. On the other hand, computation of realistic patterns of spin accumulation~\cite{Chui2008} throughout the device requires to consider balance between transport and relaxation processes. Also, the NEGF formalisms developed here, with spin-diffusion length vs selected layer thickness tuned via microscopic SO scattering terms, can tackle complicated spin pumping multilayer setups involving $I$ layers where conventional approaches are not applicable (because of spin accumulation not being well defined in an insulator).~\cite{Taniguchi2008a}

Although we do find nonzero charge current in $N|F|I|N$ multilayers when potential barrier is introduced in the device through the $I$ layer, the voltage signal $\sim 10$ pV is several orders of magnitude smaller than $\sim 1$ $\mu$V observed in experiments on such devices.~\cite{Moriyama2008} Also, this charge current is proportional to $(\hbar \omega)^2$, rather than $\hbar \omega$ for spin and charge currents in $N|F|I|F|N$ junctions or experimentally observed dc voltage signal in $N|F|I|N$ junctions.~\cite{Moriyama2008} Its origin is in asymmetry of transmission coefficients connecting the four electrodes of the dc circuit in the rotating reference frame. While the magnitude of measured voltages remains a puzzle for a variety of approaches~\cite{Xiao2008,Tserkovnyak2008,Chui2008} utilized very recently to address some of the aspects of the experiment in Ref.~\onlinecite{Moriyama2008}, we believe that combining NEGF approach to spin pumping outlined here with the density functional theory (DFT) to take into account nonequilibrium self-consistent spin and charge densities (akin to NEGF-DFT approach~\cite{Heiliger2008b} to spin-transfer torque in spin valves and MTJs) could be capable
of addressing this problem.

\begin{acknowledgments}
We thank G.~E.~W.~Bauer, T.~Moriyama, and Y.~Tserkovnyak for illuminating discussions. This work was supported by DOE Grant No. DE-FG02-07ER46374 through the Center for Spintronics and Biodetection at the University of Delaware. S.-H. Chen and C.-R. Chang also gratefully acknowledge financial support by the Republic of China National Science Council Grant No. 95-2112-M-002-044-MY3 and NSC-096-2917-I-002-127.
\end{acknowledgments}

%********************references************************************************************************

%*****************************************************************

\end{document}